\RequirePackage{fix-cm}
\documentclass[smallcondensed]{svjour3}                     

\smartqed  
\usepackage{graphicx,color}
\usepackage{amsmath}
\usepackage{mathptmx}      
\usepackage{tabularx}      
\usepackage{nicefrac}
\usepackage{natbib}
\usepackage{url}
\journalname{}
\begin{document}

\title{Pore-scale Modeling of Viscous Flow and Induced Forces in Dense Sphere Packings}

\author{Bruno Chareyre         \and
	Andrea Cortis         \and
	Emanuele Catalano         \and
	Eric Barth\'elemy
}
\institute{Bruno Chareyre \at
              Grenoble INP, UJF Grenoble I, CNRS UMR 5521, 3SR lab\\
	      BP53, 38041 Grenoble Cedex 9, France\\
              \email{bruno.chareyre@grenoble-inp.fr}           
           \and
              Andrea Cortis \at
              Earth Sciences Division \\
              Lawrence Berkeley National Laboratory \\
              Berkeley, CA, 94806, U.S.A.
           \and
              Emanuele Catalano \at
	      Grenoble INP, UJF Grenoble I, CNRS UMR 5521, 3SR lab\\
	      BP53, 38041 Grenoble Cedex 9, France
	   \and
	      Eric Barth\'elemy \at
	      Grenoble INP, UMR CNRS 5519, LEGI\\	      
	      BP 53, 38041 Grenoble Cedex 9, France
}
\date{Received: date / Accepted: date}

\maketitle

\begin{abstract}
We propose a method for effectively upscaling incompressible viscous flow in large random polydispersed sphere packings:
the emphasis of this method is on the determination of the forces applied on the solid particles by the fluid. 
Pore bodies and their connections are defined locally through a regular Delaunay triangulation of the packings.
Viscous flow equations are upscaled at the pore level, and approximated with a finite volume numerical
scheme.
We compare numerical simulations of the proposed method to detailed finite element (FEM) simulations
of the Stokes equations for assemblies of 8 to 200 spheres. 
A good agreement is found both in terms of forces exerted on the solid particles and effective permeability
coefficients.
\end{abstract}

\begin{table}
\begin{tabular}{ll}
\textbf{List of Symbols} & \\
&\\
$\alpha$ & Non dimensional conductance factor\\
$\Omega$ & Full domain of the two-phase problem\\
$\Omega_i$ & Domain defined by tetrahedron $i$\\
$\Omega_{ij}$ & union of tetrahdra ($S_{ij},P_i$) and ($S_{ij},P_j$)\\
$\Gamma$ & Part of $\Omega$ occupied by the solid phase\\
$\Gamma_i$ & Domain occupied by solid particle $i$\\
$\Theta$ & Part of $\Omega$ occupied by the fluid phase (pore space)\\
$\Theta_i$ & Part of $\Omega_i$ occupied by the fluid phase (pore)\\
$\Theta_{ij}$ & part of $\Omega_{ij}$ occupied by the fluid phase (throat)\\
$S_{ij}$ & surface of the facet $ij$, separating tetrahedra $i$ and $j$\\
$\partial X$ & contour of domain X\\
$\partial^f X$ & part of contour of X intersecting the fluid phase\\
$\partial^s X$ & part of contour of X intersecting (or in contact with) the solid phase\\
$\gamma_{ij}$ & area of $\partial\Theta_{ij}$ in contact with spheres\\
$\gamma_{ij}^k$ & area of the part of $\partial\Theta_{ij}$ in contact with sphere $k$\\
$S_{ij}$ & the common facets of tetrahedra $\Omega_i$ and $\Omega_j$\\
$A^f_{ij}$ & area of the fluid part $S_{ij} \cap \Theta$ of facet $ij$\\
$A^k_{ij}$ & area of the intersection $S_{ij} \cap \Gamma_k$ of facet $ij$ and sphere $k$\\
$P_i$ & Voronoi dual (weighted center) of tetrahedra $i$\\
$p'$ & microscopic (pore-scale) fluid pressure\\
$p_i$ & macroscopic fluid pressure in tetrahedra $i$\\
$\vec{u}'$ & microscopic fluid velocity\\
$\vec{u}$ & macroscopic fluid velocity\\
$\vec{v}$ & geometric contour velocity\\
$q_{ij}$ & flux through facet $ij$\\
$V^f_i$ & fluid volume contained in pore $i$\\
$R^h_{ij}$ & hydraulic radius of throat $ij$\\
$R^{eff}_{ij}$ & effective radius of throat $ij$\\
$\mu$ & dynamic viscosity\\
$L_{ij}$ & length of throat $ij$\\
$F_{x}^y$ & forces exerted by the fluid on the solid phase, $x$ and $y$ denote different terms in forces decomposition\\
$g_{ij}$ & hydraulic conductance of facet (throat) $ij$\\
$K_{ij}$ & hydraulic conductivity of facet (throat) $ij$
\end{tabular}
\end{table}

%
\section{Introduction}
\label{sec:intro}

Understanding the mechanical behavior of fluid-saturated granular systems (e.g., soils, rocks,
concretes, powders, and chemical reactors) hinges upon a detailed description of the stress exchange
between fluid and solid phases. Fluid flowing through a matrix of solid spheres, exerts forces on the
solid grains, displacing them from the trajectory they would have had if left to interact alone in a dry medium.
In turn, the displacement of the grains creates a dynamically changing domain for the fluid flow. The combination of
these two effects results in nontrivial physically measurable effects, which are hard to capture by
first-principles derivations alone. Despite the important practical applications that are touched by fluid-saturated granular systems, no model currently
exists that can efficiently reproduce the precise evolution of large systems over different stress conditions in 3D problems. 

This work represents a first step in the direction of developing a fully-coupled, computationally efficient model for
the evolution of fluid-saturated granular materials under stress. In particular, we will focus our effort on the
faithful approximation of the forces applied by the fluid on the solid grains, with the aim of incorporating these
forces in the  \textit{discrete element method} (DEM) computations \citealt{Cundall1979}.

While the movement of the solid grains can be efficiently modeled by DEM computations, viscous fluid computations in
dynamically changing domains present incredible computational challenges. At the microscopic (sub-pore) scale, fluid
flow is governed by Stokes equations, which express fluid mass and momentum conservation at small Reynolds and Stokes
numbers:
\begin{gather}
\nabla p = \mu \, \nabla^2 \vec{u} - \rho\nabla\Phi , \label{eq:stokes} \\
\nabla \cdot \vec{u} = 0
\label{eq:cont}
\end{gather}
where $\vec{u}$, and $p$ are the microscopic fluid velocity and piezometric pressure, respectively, $\mu$
is the fluid dynamic viscosity, and $\Phi$ is a potential field (e.g., gravitational field). The piezometric
pressure $p$ is related to the absolute pressure $p*$ via $p=p*-\rho\Phi$
Equations \eqref{eq:stokes}-\eqref{eq:cont} are augmented with a no-slip boundary condition for the fluid velocity,
 at the grain boundaries, $\vec{u} = 0$, which is essentially responsible for the microscopic viscous energy losses
(drag) that translate in a net loss of the macroscopic piezometric pressure, $p$, over the length of a porous column.

At centimeter scale (and above), for relatively homogeneous porous materials, fluid flow is governed by Darcy law, which
states the linear proportionality between the macroscopic pressure gradient $\nabla p$ and the fluid flux $q$
(discharge per unit area) :
\begin{equation}
q = -\frac{k}{\mu} \, \nabla p,
\label{eq:darcy}
\end{equation}
where $k$ is the permeability. Darcy law emerges as a volume averaging of the viscous forces applied by the
fluid flow at the surface of the grains over a sufficiently large number of pores.

Recent advances in porous media imaging techniques give access to an unprecedented level of pore-space detail,
down to the micrometer scale and below, where the Stokes equations could be, in principle, solved numerically.
The numerical solution of the Stokes equations in spheres assemblies is, however, computationally expensive, 
and becomes prohibitive on commodity desktop computers for number of particles exceeding the hundreds.
On the other hand, typical DEM simulations, can easily handle 
assemblies with number of particles ranging between $10^4$ and $10^5$ on the same commodity desktop computers, and compute their evolution over $10^5$ time-steps or more 
(see e.g. \citealt{Scholtes2009}).
It is therefore obvious that a direct coupling of Stokes equations and DEM simulations becomes unfeasible for
real-world problems, hence the need to develop an upscaled fluid model that minimizes the ratio between the degrees of
freedom (DOFs) in the fluid and the number of solid particles.

The past thirty years have seen a considerable effort in the exact upscaling of Equations
\eqref{eq:stokes}-\eqref{eq:cont}, i.e., on how to obtain expressions for the permeability $k$ from the knowledge of the
microscopic
flow. Upscaling techniques such as the Volume Averaging with Closure (\citealt{whitaker1999method}),
and homogenization (\citealt{sanchez1985homogenization}) invariably require the solution of a
(tensorial) Stokes problem, which has essentially the same algorithmic complexity of the
``original'' Stokes fluid flow problem. 
In this work we will not be concerned with this type of theoretical upscaling issues. Rather, we will focus our
attention to the derivation of a physically based, simplified model of flow through porous materials that overcomes the
numerical issues associated with the solution of the full Stokes problem.

Different strategies have been adopted in the past for the solution of the fluid-solid coupling, which essentially
differ in the modeling techniques adopted for the fluid part of the problem, as reviewed below.

\paragraph{Microscale Stokes flow modeling\\}
The numerical solution of the Stokes equations is a computationally demanding task, especially
for complex three-dimensional pore geometries. Finite Element Methods (FEMs) are often used because of their flexibility
in the definition of the numerical mesh. 
FEM meshes in three dimensions tend to be, however, very large and so do the computer memory footprint and
computational times needed for the solution of the associated nonlinear system of equations. 
\cite{Glowinski2001}, for instance, reported computational costs of approximately 0.1s per particle and per time-step in
fluidization simulations on a SGI Origin 2000, employing an advanced fictitious domain method and a partially
parallelized code: extrapolating these values to a number of particle typical of DEM problems, results in computational
times in the order of 3-30 years.

A numerical method that does not resort to the solution of large systems of nonlinear equations is the
Lattice-Boltzmann (LB) scheme. While LB is generally faster than the FEM and has the possibility of being easily
parallelized on multicore machines and GPUs (\url{http://sailfish.us.edu.pl/}), commonly implemented fixed size grids
can result in considerably larger computer memory occupancy in three dimensions.
Only recently, grid refinement schemes have started to be implemented in open source LB codes
(\url{http://www.lbmethod.org/palabos/}). Realistic 
microscale flow simulations in complex pore geometries still require, however, access to large CPU clusters.

\paragraph{Continuum-discrete Darcy flow modeling\\}
In order to get acceptable computational costs, a number of authors considered coarse-grid CFD methods
(\citealt{NakasaBook,Kawaguchi1998,McNamara2000,Kafui2002,Zeghal2004,Niebling2010}). In such methods, flow and
solid-fluid
interactions are defined with simplified semi-empirical models based on Darcy law. There is no direct coupling at the
local scale: the forces acting on the individual particles are defined as a function of mesoscale-averaged fluid
velocity obtained from porosity-based estimates of the permeability. 
Such strategies have been adopted for dilute particles suspensions as well as for dense materials at low strain rate
(e.g., soils, rocks). Continuum-discrete couplings succeed in making coupled problems affordable in terms of CPU cost.
The use of phenomenological laws for the estimation of the permeability, however, limits severely the predictive power
of these models in uncalibrated parameter regions. Ultimately, such strategies do not render correctly the individual
particle behavior, and cannot be reliably applied to problems such as strain localization, segregating phenomena, effects of local heterogeneities in porosity, and internal
erosion by transport of fines.

\paragraph{Pore-network modeling\\}
Pore-network modeling covers a third class of models, based on a simplified representation of porous media as a network
of pores and throats. Pore-network models have been most commonly developed to predict the permeability of materials from microstructural geometry
(\citealt{Bryant1993,Thompson1997,Bakke1997,Patzek2001,Hilpert2003,Abichou2004}), but have also been extended to include multiphase flow effects (e.g., air bubbles, immiscible two-phase flow) (\citealt{Bryant1992,Oren1998,Bryant2003,Piri2005}). Crucial for their success is an adequate definition of how fluids are exchanged between
pores
in term of the local pore geometry. This aspect will be discussed further in section \ref{sec:transmissivity}.

\paragraph{Deformable pore-network modeling\\}
Pore-network models studies have mostly focused on flow in passive rigid solid frames. 
Little attention, however, has been devoted to the definition of forces applied to individual
particles in the solid phase. The main aim of this work is to develop expressions for these forces,  
to be used in three-dimensional DEM simulations of solid particles immersed in a fluid flow.

Early ideas of a coupled pore-network flow and DEM can be found in the works of Hakuno and Tarumi
\citeyearpar{HakunoBook}
and later by Bonilla \citeyearpar{Bonilla2004}. These studies, however, were limited to 2D models of
discs assemblies, where pores were defined by closed loops of particles in contact. Since such pore geometry does
not offer any free path for fluid exchanges, 2D problems implied some arbitrary definition of the local conductivity,
assuming virtual channels between adjacent voids. Adapting this approach to 3D spheres assemblies 
enables the definition of the local hydraulic conductivity using the actual geometry of the packing, as spheres packings
always define an open network of connected voids. This in turns opens up the possibility of predicting both the macro-scale
permeability and the forces acting on the individual particles rather than postulating it: this is the central theme of this work.
%
\section{Upscaled fluid flow model}
\label{sec:FVformulation}
As described in the previous section, fine-scale Stokes flow numerical simulations present a prohibitive computational cost and an effective,
upscaled fluid flow model is needed. In this section, we present a detailed derivation of the
proposed upscaled fluid flow model, starting from the decomposition of the pore space in terms of pore bodies and local
conductances. This decomposition results naturally in a finite volume description of the flow equations. The
expressions for the forces on the individual particles are then derived. Finally, we describe the actual implementation
of the model.

\subsection{Pore-scale volume decomposition}
\label{sec:volume_decomposition}
\begin{figure}
\centering
\includegraphics[scale=1.2]{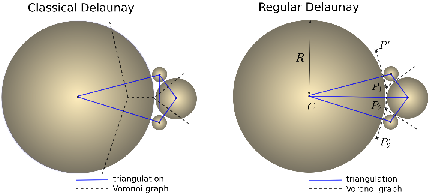}
\caption{Comparison of 2D triangulations and dual Voronoi graphs obtained from classical Delaunay and regular Delaunay:
classical Voronoi graph has branches inside discs, while regular Voronoi gives all branches in the voids space. Observe
that if R tends to infinity, \{$P'_1$,$P_1$,$P_2$,$P'_2$\} tends to aligned points.}
\label{fig:reg_vs_nonreg}
\end{figure}

Delaunay triangulations and their dual Voronoi graphs are widely used for structural studies of molecules, liquids,
colloids, and granular materials (\citealt{Aurenhammer91}). More specifically, it has been applied to domain
decompositions
in sphere packings, for the definition of microscale strains and stresses (\citealt{Calvetti1997,Bagi2006,Jerier2010}),
and
pore-scale modeling of single-phase or multi-phase flow (\citealt{Bryant1992,Bryant2003,Thompson1997}). The most common
type of Delaunay triangulation (hereafter referred to as \textit{classical Delaunay}) is defined for a set of
isolated points. In this case the dual \textit{classical Voronoi} graph of the triangulation defines polyhedra enclosing
each point in the set, and facets of the polyhedra are parts of planes equidistant to two adjacent points. For
mono-sized spheres, this construction gives a map of the void space. Classical Delaunay-Voronoi graphs, however, present a number
of misfeatures when applied to spheres of different sizes. Notably, branches and facets of the Voronoi graph can cross non-void regions, or divide
regions in such a way that the splits are orthogonal to the actual void, as illustrated in fig.~\ref{fig:reg_vs_nonreg}.

\citealt{Luchnikov1999} proposed to define the split between spheres by means of curved surfaces equidistant from both spheres
surfaces, thus defining \textit{Voronoi S} cells. The use of Voronoi S graphs is, however, computationally expensive for large
number of particles. \textit{Regular Delaunay} triangulation (see fig.~\ref{fig:reg_vs_nonreg}), while
overcoming the problems associated with classical Delaunay triangulations, provides a computationally cheap alternative
to the use of Voronoi S cells.

\begin{figure}
\centering
\includegraphics[width=75mm]{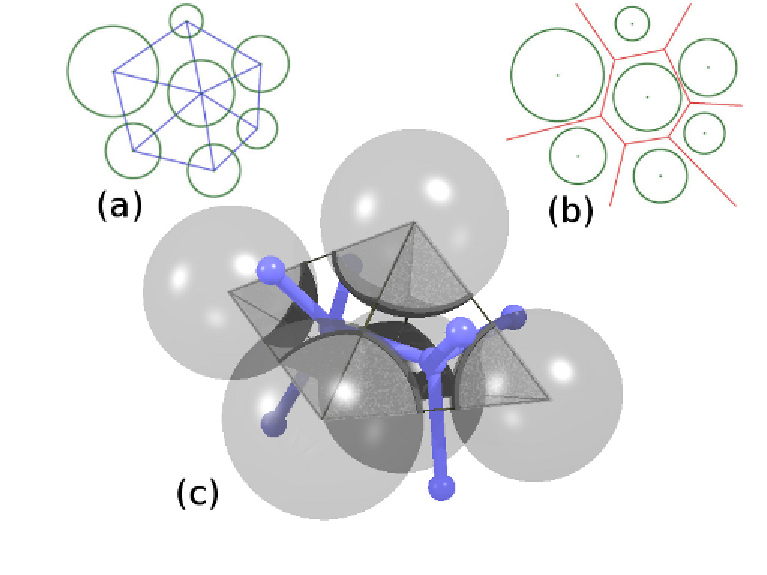}
\caption{Adjacent tetrahedra in the regular Delaunay triangulation and dual Voronoi network, in two dimensions (a,b) and
three dimensions (c).}
\label{fig:delaunay_cells}
\end{figure}

Regular Delaunay triangulation (\citealt{Edelsbrunner1996}) generalizes classical Delaunay triangulation to
weighted points, where weights account for the radius of spheres. It can be shown that the dual Voronoi graph of the regular
triangulation is entirely contained in voids between spheres, as opposed to the dual of classical Delaunay, as
seen on fig. \ref{fig:reg_vs_nonreg} (2D) and fig. \ref{fig:delaunay_cells} (2D/3D). 
Edges and facets of the regular Voronoi graph are lines and planes, thus enabling fast computation of geometrical
quantities. In section \ref{sec:transmissivity} we will use a combination of regular
Delaunay facets and regular Voronoi vertices to decompose the pore volume.

In what follows, $\Omega$ denotes a domain occupied by a porous material, in which $\Gamma$ and $\Theta$ are the domains
occupied respectively by the solid and the fluid: $\Omega=\Gamma \cup \Theta$, $\Gamma \cap \Theta = \emptyset$
($\Theta$ is also called \textit{pore space}). 
We denote by $N_c$ the number of tetrahedral cells in the regular Delaunay
triangulation of the sphere packing, and $\Omega_i$ the domain defined by tetrahedron $i$~: 
$\Omega=\cup _{i=1}^{N_c}{\Omega_i}$. Similarly, $N_s$ is the number of spheres, and $\Gamma_i$ the domain
occupied by sphere $i$, so that $\Gamma=\cup _{i=1}^{N_s}{\Gamma_i}$.

%
\subsection{Fluxes}
\label{sec:goveq}
\subsubsection{Continuity}
\label{sec:continuityeq}
\begin{figure}
\centering
\includegraphics[width=65mm]{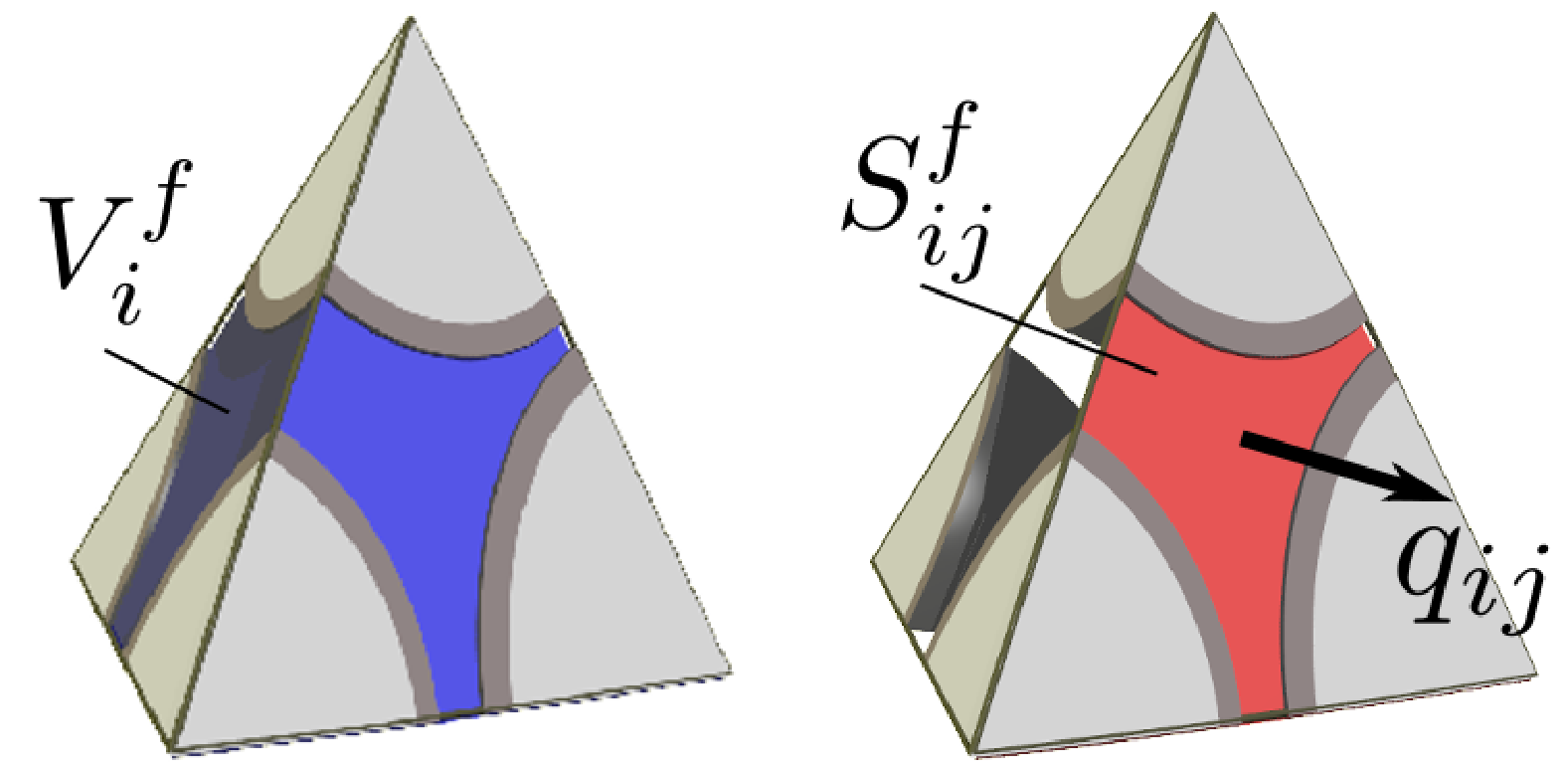}
\caption{Tetrahedral element of the finite volume decomposition.}
\label{fig:tetrahedron}
\end{figure}

We define $\Theta_i$ as the portion of the tetrahedral element $\Omega_i$ not occupied
by spheres. The volume $V_i^f$ of $\Theta_i$ is filled by fluid (fig. \ref{fig:tetrahedron}), since we are considering saturated porous media.
The continuity equation for an incompressible fluid \eqref{eq:cont}, cast into its surface integral form and using the
divergence theorem, gives a relation between the time derivative of $V_i^f$ and the fluid velocity:
\begin{equation}
\dot{V_i^f} = \int\limits_{\partial\Theta_{i}} (\vec{v}-\vec{u}) \cdot \vec{n} \ \text{d}s ,
\end{equation}
where $\partial\Theta_{i}$ is the contour of $\Theta_{i}$, $\vec{n}$ the outward pointing unit vector normal to
$\partial\Theta_{i}$, $\vec{u}$ the fluid velocity, and $\vec{v}$ the velocity of the contour. One part of the contour,
noted $\partial^s\Theta_{i}$, corresponds to a solid-fluid interface. At any point in $\partial^s\Theta_{i}$, we have
$(\vec{v}-\vec{u})\cdot\vec{n}=0$, so that the integration domain above can be restricted to $\partial^f\Theta_{i}$, the
fluid part of the contour. By introducing $S^f_{ij} (j \in\{j_1,j_2,j_3,j_4\}$ the intersections of triangular surfaces
$S_{ij}$ with the fluid domain (fig. \ref{fig:tetrahedron}), so that $\partial^f\Theta_{i}=\cup
_{j=j_1}^{j_4}{S^f_{ij}}$, we can define four intergrals describing fluid fluxes $q_{ij}$ from tetrahedron $i$
to adjacent tetrahedra $j_1$ to $j_4$. Finally: 
\begin{equation}
\dot{V_i^f} = \sum\limits_{j=j_1}^{j_4}\int\limits_{S^f_{ij}}(\vec{v}-\vec{u}) \cdot \vec{n} \ \text{d}s = -
\sum\limits_{j=j_1}^{j_4}q_{ij}
\end{equation}
In deformable sphere packings, $\dot{V_i^f}$ can be computed on the basis of particles motion (i.e. velocities of the vertices of the tetrahedra), thus linking fluid fluxes with the deformation of the solid skeleton. The derivation of $\dot{V_i^f}$ as
function of particles velocities is not detailed here for brevity, because in what follows particles are assumed to be
fixed: $V_i^f$ is constant and the continuity equation simplifies to 
\begin{equation}
\sum\limits_{j=j_1}^{j_4}q_{ij} = 0 .
\label{eq:continuityeq}
\end{equation}

%
\subsubsection{Local conductance}
\label{sec:transmissivity}
Both Stokes (\ref{eq:stokes}) and Darcy (\ref{eq:darcy}) equations imply a linear relation between pressure gradients
and fluxes. Here, we introduce a \textit{inter-pore} gradient, defined as the ratio between pressure difference
$p_i-p_j$ between two connected tetraheral cells, 
and a relevant length $L_{ij}$ - to be defined below. 
Being linear, the relation between $q_{ij}$ and the inter-pore gradient
can be expressed using the \textit{local conductance} factor $g_{ij}$ of facet $ij$:
\begin{equation}
q_{ij} = g_{ij} \ \frac{p_i-p_j}{L_{ij}} .
\label{eq:flux}
\end{equation}
Considerable efforts have been devoted in the past literature to the definition
of $g_{ij}$ in pore-network models, with attempts to generalize Poiseuilles's law to pores of different shapes,
eventually mapping the microstructure of some real materials for permeability predictions
(\citealt{Piri2005,Hilpert2003}).
By analogy with the Hagen-Poiseuille relation, $g_{ij}$ may be defined by introducing the
\textit{hydraulic radius} of the pore throat $R^h_{ij}$, and its cross-sectional area $A_{ij}$ (the definitions of
$R_{ij}$ and $A_{ij}$ are discussed below), by means of a non-dimensional conductance factor $\alpha$ reflecting the
throat's shape (Hagen-Poiseuille for circular cross-sections of radius $2R^h_{ij}$ is recovered with
$\alpha=\nicefrac{1}{2}$):
\begin{equation}
g_{ij} = \alpha \ \frac{{A_{ij}\,R^h_{ij}}^2}{\mu} .
\label{eq:poiseuillecylinder} 
\end{equation}

$L_{ij}$, $R^h_{ij}$, and $A_{ij}$ are geometrical quantities describing the throats geometry. 
Even though these variables are found in most of the papers cited above, there is no general agreement
on their definition for pore-network modeling of arbitrary microstructural geometry. A detailed analysis of the effect of non-circular
cross-sections with variable constriction along the flow path can be found in \citealt{Patzek2001} and \cite{Mortensen2005},
while \cite{Thompson1997},\cite{Bakke1997}, and \cite{Bryant1993} focused specifically on  triangulated sphere packings.
Some of these models, however, do not clearly define a partition of the void space, and as such they need ad-hoc corrections for $L_{ij}$ to ensure that the same pore volumes are not accounted for multiple times in different cells (\citealt{Bryant1993}).
We observe here that in \textit{classical} Delaunay-Voronoi graphs of spheres with polydispersed radii, circumcenters (and barycenters) may lie inside the solid phase (\cite{Bryant1992}), and definitions of $L_{ij}$ based on distances between circumcenters become problematic.

\begin{figure}
\centering
\includegraphics[width=45mm]{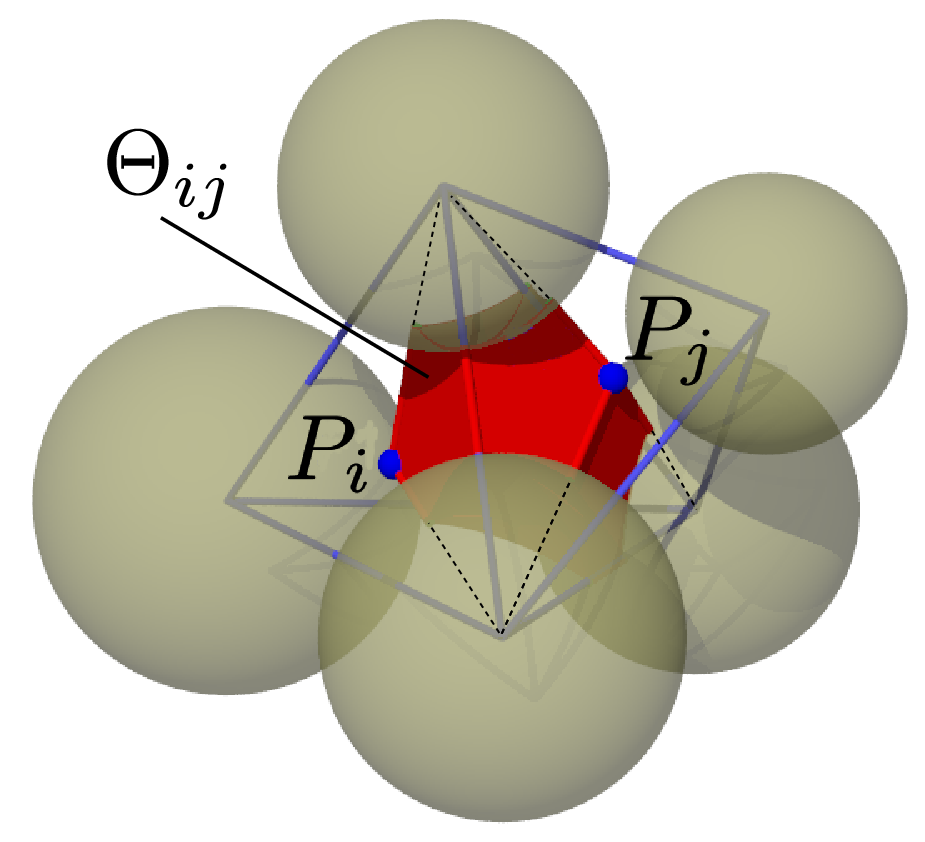}
\includegraphics[width=75mm]{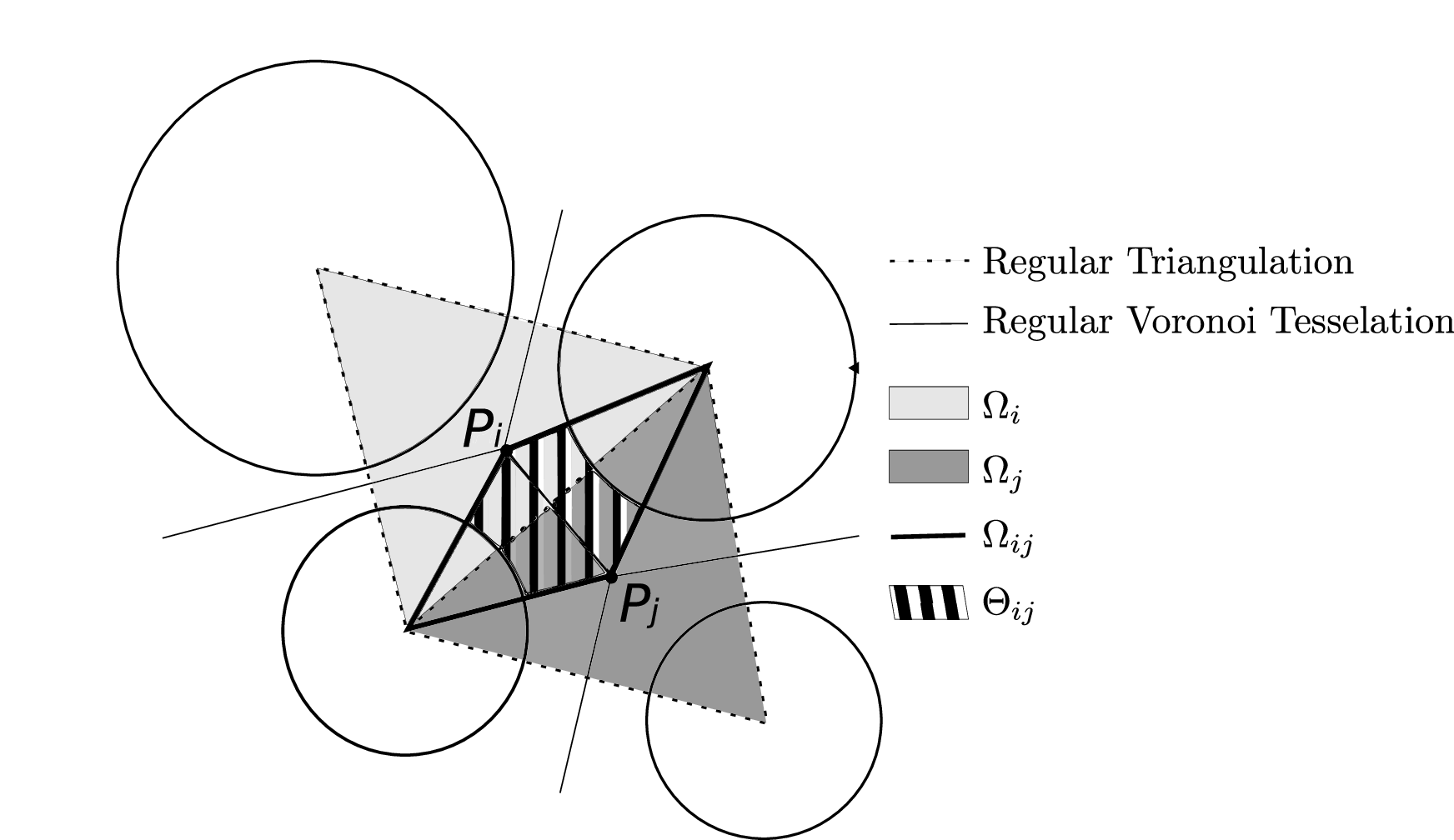}
\caption{Construction of subdomain $\Theta_{ij}$, defining the volume of the throat assigned to a facet for the
definition of hydraulic radius $R_{ij}^h$; in 3D (left) and 2D (right).}
\label{fig:teta_ij}
\end{figure}

\begin{figure}[!ht]
\centering
\includegraphics[width=95mm]{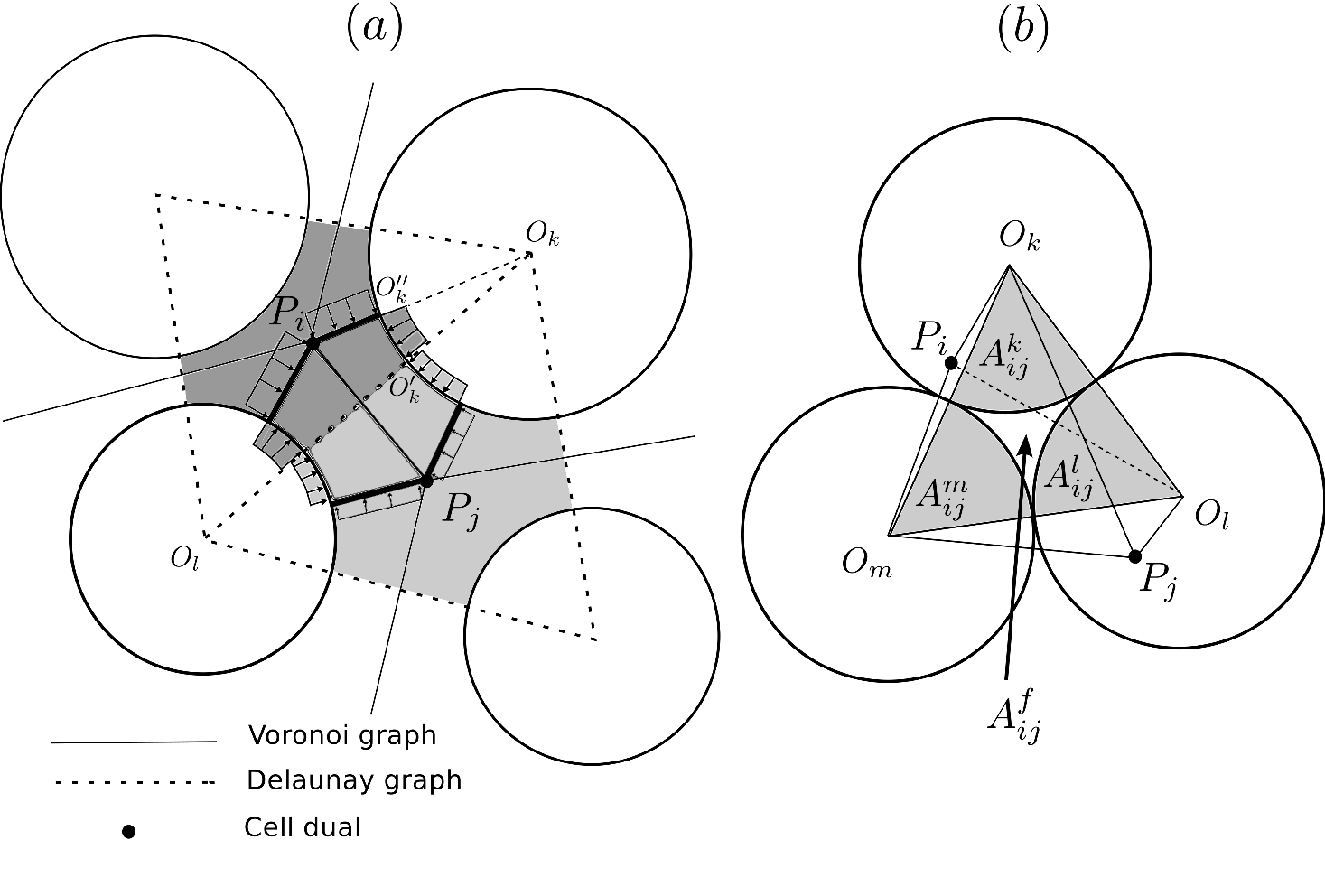}
\caption{Volume decomposition for viscous drag force definition : (a) pressure distribution on $\partial\Theta_{ij}$ (in
2D for clarity), (b) definition of facet-spheres intersections in 3D.}
\label{fig:forcescheme}
\end{figure}

For the present model, we take advantage of the regular Voronoi graph structure, whose edges and vertices are always
contained in the pore space. A partition of the full domain is defined as the set of sub-domains $\Omega_{ij}$,
with $\Omega_{ij}$ the union of two tetrahedra constructed from facet $S_{ij}$ and Voronoi vertices $P_i$ and $P_j$ :
$\Omega_{ij}=(S_{ij},P_i)\cup(S_{ij},P_j)$.
Keeping only the part of $\Omega_{ij}$ that intersect the pore space, we define a throat connecting two pores, noted
$\Theta_{ij}$ (see figs. \ref{fig:teta_ij} and \ref{fig:forcescheme}). We define the hydraulic radius $R^h_{ij}$ in eq. \ref{eq:poiseuillecylinder} as the ratio between the throat's volume
and solid-fluid interfaces area. 
Denoting by $\phi_{ij}$ the volume of $\Theta_{ij}$, and $\gamma_{ij}$ the area of $\partial^s\Theta_{ij}$ (the part of the
contour in contact with spheres), the \textit{hydraulic radius} $R^h_{ij}$ reads:
\begin{equation}
R_{ij}^h=\phi_{ij}/\gamma_{ij}
\label{eq:hydraulicradius} 
\end{equation}
Note that, with the partition of the voids space proposed here, the definition of effective radii accounts for the exact
total solid surface and
void volume in the packing. Noting $\cal F$ the set of facets, we have $
\Theta=\bigcup_{ij\in\cal F}{\Theta_{ij}}$, and $\partial\Theta=\bigcup_{ij\in\cal F}{\partial^s\Theta_{ij}}$.

We define the length of the throat as the distance between Voronoi vertices: $L_{ij}=\lVert P_{i}P_{j}\rVert$, and the
throat's cross-sectional area as that of the fluid domain $S_{ij}^f$ introduced in the previous section (fig.
\ref{fig:tetrahedron}).
In this model, the factor $\alpha$ is uniquely assigned to all
throats of a packing and can be considered a calibration
factor. A value of $\alpha=\nicefrac{1}{2}$ has been choosen as
an initial guess, by analogy with the Hagen-Poiseuille law. The numerical simulations presented in section \ref{sec:comparisons} indicate that setting alpha=1/2 is indeed appropriate for the spheres geometries considered in this investigation. We observe that the values of $\alpha$ obtained analytically vary in a
moderately wide range of values for compact throats sections (\citealt{Patzek2001}). It is also worth noting that, the
factor $\alpha$ being
the same for all throats, modifying this value will not modify the pressure distribution in the boundary value problems
presented below
(mixed Neumann-Dirichlet), and consequently it would not affect the values of forces applied on the particles.

Another possible definition of $g_{ij}$ has been proposed by \cite{Bryant1992} for dense packings of mono-sized spheres.
These authors defined $g_{ij}$ as a function of an \textit{effective radius} $R_{ij}^{eff}$ (ER), accounting thus only for the effects of the narrowest cross-section (surface $S_{ij}$ in
fig.~\ref{fig:tetrahedron}) and disregarding the influence of the full volume of the flow path. The effective radius is defined
\footnote{For the convenience of comparisons, and consistently with the expression of conductance we are using, we
introduce the effective radius $R^{eff}$ as the half of its definition in the original paper from Bryant and Blunt. The
final value of the conductance is preserved.}
as $R_{ij}^{eff}=(r_{ij}^{eq.}+r_{ij}^{inscr.})/4$, where $r_{ij}^{eq.}$
is the radius of the disk of same surface as $S_{ij}^f$, and $r_{ij}^{inscr.}$ is the radius of the circle inscribed in
the three spheres intersecting the facet. It represents the hydraulic radius of the circular tube that would have the
same conductance as the throat: 
\begin{gather}
g_{ij} = \frac{{2\,\pi\,R^{eff}_{ij}}^4}{\mu} = \frac{{A^{eff}_{ij}}\,{R^{eff}_{ij}}^2}{2\mu} ,
\label{eq:Reffg} 
\end{gather}
where $A^{eff}_{ij}=\pi (2 R^{eff}_{ij})^2$ is the cross-sectional area of the equivalent tube. The second expression is
given for the purpose of comparison with eq. \ref{eq:poiseuillecylinder}.
The values of $R_{ij}^{h}$ and $R_{ij}^{eff}$ in random dense packings of polydispersed spheres, and the results they
give in terms of fluxes and forces are compared in section \ref{sec:comparisons}.

%
\subsection{Forces on solid particles}
\label{sec:forces_formulation}

The total force $\vec{F}^{k}$ generated on particle $k$ by the fluid includes the effects of absolute pressure $p^*$ and
viscous stress $\vec{\tau}$, which divergence is the right-hand side of \eqref{eq:stokes}:
\begin{equation}
\vec{F}^k=\int\limits_{\partial\Gamma_{k}}{(p^*\, \vec{n}+\vec{\tau \, n}) \ \text{d}s} .
\end{equation}
We recall that the piezometric pressure governing the flow problem was defined as $p=p^*-\rho\Phi(\vec{x})$. By
introducing $p$ in the previous equation,
 one can define $\vec{F}^k$ as the sum of three terms, noted $\vec{F}^{b,k}$, $\vec{F}^{p,k}$ and $\vec{F}^{v,k}$:
\begin{equation}
\vec{F}^k=\int\limits_{\partial\Gamma_{k}}{\rho\Phi(\vec{x})\,\vec{n} \text{d}s}+\int\limits_{\partial\Gamma_{k}}{p\,
\vec{n} \text{d}s}+\int\limits_{\partial\Gamma_{k}}{\vec{\tau \,
n}\,\text{d}s}=\vec{F}^{b,k}+\vec{F}^{p,k}+\vec{F}^{v,k} .
\label{eq:forceterms2}
\end{equation}
$\vec{F}^{b,k}$ is the so-called \textit{buoyancy} force, and can be computed independently. In the case of
gravitational body forces, it will give the Archimede's force proportional to the volume of $\Gamma_k$. $\vec{F}^{p,k}$
and $\vec{F}^{v,k}$ are terms resulting from viscous flow: $\vec{F}^{p,k}$ resulting from losses of piezometric
pressure, and $\vec{F}^{v,k}$ resulting from viscous shear stress.
Both $\vec{F}^{p,k}$ and $\vec{F}^{v,k}$ will be derived separately, at the scale of the domains $\Omega_{ij}$ that were
introduced in the previous section, in the approximation of piecewise constant pressure.

\subsubsection{Integration of pressure}
The force generated by $p$ on particle $k$ in domain $\Omega_{ij}$ is the sum of two terms implying pressures $p_i$ and
$p_j$ (see fig.~\ref{fig:forcescheme}),
\begin{equation}
\vec{F}^{p,\, k}_{ij} = \int\limits_{\partial\Gamma_{k}\cap \Omega_{i}\cap\Omega_{ij}}p_i \, \vec{n} \, \text{d}s +
\int\limits_{\partial\Gamma_{k}\cap \Omega_{j}\cap\Omega_{ij}}p_j \, \vec{n} \, \text{d}s 
\label{eq:buoyancyforce}
\end{equation}
In three dimensions, computing such integral on spherical triangles could be computationally
expensive. It is more convenient to project the pressure on the conjugate planar parts (angular sectors) of the closed
domain
$\Gamma_k\cap\partial\Omega_{ij}$, which trace in the plane of fig.~\ref{fig:forcescheme} corresponds to segments $OO'$
and $OO''$.
Note that when iterating over all domains $\Omega_{ij}$ adjacent to one particle, the integral on sectors of type $OO''$
will appear twice
with opposite normals and cancel out each other. Finally, the
contribution to pressure force on particle $k$ in domain $\Omega_{ij}$ is simply proportional to the area $A^k_{ij}$ of
sector $S_{ij}\cap\Gamma_k$ (trace $OO'$). If $\vec{n}_{ij}$ is the unit vector pointing from $P_i$ to $P_j$, the force
reads:
\begin{equation}
\vec{F}^{p,k}_{ij} = A^k_{ij} \ (p_{i}-p_{j}) \ \vec{n}_{ij}
\label{eq:pressureforcek}
\end{equation}

\subsubsection{Integration of viscous stresses}
To integrate the viscous stresses, we first define the total viscous force $\vec{F}^v_{ij}$
applied on the solid phase in $\Omega_{ij}$. Since $\Omega_{ij}$ intersects three spheres, $\vec{F}^v_{ij}$ will have to
be later splited into three terms. $\vec{F}^v_{ij}$ is defined as
\begin{equation}
\vec{F}^v_{ij}=\int\limits_{\partial^s\Theta_{ij}}{\vec{\tau\,n} \ \text{d}s}.
\label{eq:viscousij}
\end{equation}

An expression of $\vec{F}^v_{ij}$ is obtained by integrating the momentum conservation equation \eqref{eq:stokes}) in
$\Theta_{ij}$. The integral is cast in the form of a surface integral on contour $\partial\Theta_{ij}$ using the
divergence theorem :
\begin{equation*}
\int\limits_{\partial\Theta_{ij}}{(p \, \vec{n}+\vec{\tau \, n}) \ \text{d}s} = 0 .
\end{equation*}
This integral can be decomposed as in eq. \ref{eq:momentum}, where the terms correspond respectively to
$\vec{F}^v_{ij}$, to the sum of viscous stress on the fluid part $\partial^f\Theta_{ij}$ of the contour, and to the sum
of pressure. By neglecting the second term (thus assuming that pressure gradients are equilibrated mainly by viscous
stress on the solid phase), the expression of $\vec{F}^v_{ij}$ takes the form of equation \ref{eq:dragforce}, where the
viscous force is simply proportional to the product of the throat's cross-sectional area $A^f_{ij}$ and the pressure
jump $p_j-p_i$.
\begin{equation}
\int\limits_{\partial^s\Theta_{ij}}{\vec{\tau \, n}\,\text{d}s}+\int\limits_{\partial^f\Theta_{ij}}{\vec{\tau \,
n}\,\text{d}s}+
\int\limits_{\partial\Theta_{ij}}{p\,\vec{n}\,\text{d}s} = 0 .
\label{eq:momentum}
\end{equation}

\begin{equation}
\vec{F}^v_{ij}
\simeq -\int\limits_{\partial\Theta_{ij}}{p \, \vec{n} \ \text{d}s} \ = \ A^f_{ij} \, (p_j-p_i) \, \vec{n}_{ij}
\label{eq:dragforce}
\end{equation}

In order to define the viscous forces applied on each of the three spheres intersecting $\Omega_{ij}$, it is assumed
that the force on sphere $k$ is proportional to the surface of that sphere contained in the subdomain. If
$\gamma^k_{ij}$ denotes the area of the curved surface $\partial\Gamma_k\,\cap\Omega_{ij}$, the force on $k$ then reads:
\begin{equation}
\vec{F}^{v,\, k}_{ij}=\vec{F}^v_{ij} \ \frac{\gamma^k_{ij}}{\sum_{k=1}^3\gamma^k_{ij}}
\label{eq:dragforcek}
\end{equation}
Finally, the total force on one particle is obtained by summing viscous and pressure forces from all incident facets
with the buoyancy force:

\begin{equation}
\vec{F}^k = \sum\limits_{(ij)_{incident}}\{\vec{F}^{v,k}_{ij} + \vec{F}^{p,k}_{ij}\}+\vec{F}^{b,k}
\label{eq:totalforce}
\end{equation}

\subsection{Implementation}
\subsubsection{Network definition}
The network model has been implemented in C++, and it is freely available as an optional package of the open-source
software Yade (\citealt{YadeDoc}). The C++ library CGAL~(\citealt{Boissonnat2002}) is used for the triangulation
procedure.
This library ensures exact predicates and constructions, and is one of the fastest computational geometry codes
available (\citealt{Liu2005}). 
It is worth noting that regular triangulation involves only squared distances comparisons, thus avoiding time consuming
square roots and divisions. The computation of forces (eq. \ref{eq:totalforce}) also requires only simple vector
multiplications. The only non-trivial operation is the
computation of solid angles needed to define the volumes and surfaces associated to sphere-tetrahedron intersections, in
eq.\ref{eq:hydraulicradius}. This cost can be reduced significantly, however, using the algorithm of Oosterom and
Strackee~\citeyearpar{Oosterom1983}.

\subsubsection{Flux problem solution}
Combining Equations \ref{eq:continuityeq} and \ref{eq:flux} gives for element $i$:
\begin{equation}
\sum\limits_{j=1,..,4}{\frac{g_{ij}}{L_{ij}} \, (p_i-p_j)}=\sum\limits_{j=1,..,4}{K_{ij} \, (p_i-p_j)}=\dot{V^f_i}
\label{eq:continuityeq_discrete}
\end{equation}

If the triangulation defines $N_c$ tetrahedra and if the number of elements with imposed pressure is $N_p$,
the pressure field is obtained after solving a linear system of size $N_c-N_p$. The matrix of the system is sparse and
symmetric ($K_{ij}=K_{ji}$), and positive defined. An over-relaxed Gauss-Siedel algorithm has been used
to solve this linear system in the simulations presented below. This algorithm has been chosen for its simplicity,
though other strategies could be preferred in the future for better performance and for complex couplings (including
particles motion). 
%
\section{Comparison with small-scale FEM simulations}
\label{sec:comparisons}

\subsection{Numerical setup}
In this section, we compare the results of the pore-scale finite volume (FV) formulation described in section
\ref{sec:FVformulation}
with FEM Stokes flow simulations of Equ. \eqref{eq:stokes}-\eqref{eq:cont}, for dense sphere packings subjected to
an imposed pressure gradient.
We consider sphere packings contained in a cube of size $l_0$ (all results below will be normalized with respect to this reference length). Boundaries are accounted for in the
triangulation process in the form of spheres with near-infinity radii ($10^6\times l_0$).
Hence, planes are not introducing special cases to handle in the algorithms, and equations presented in
previous section apply for boundaries as for any other sphere.

\begin{figure}
\centering
\includegraphics[width=65mm]{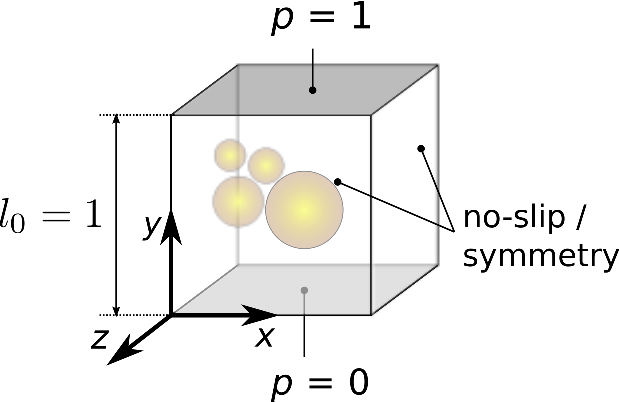}{(a)}
\includegraphics[width=45mm]{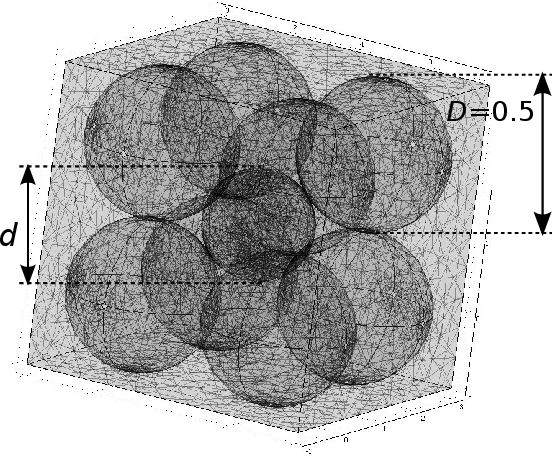}{(b)}
\includegraphics[width=45mm]{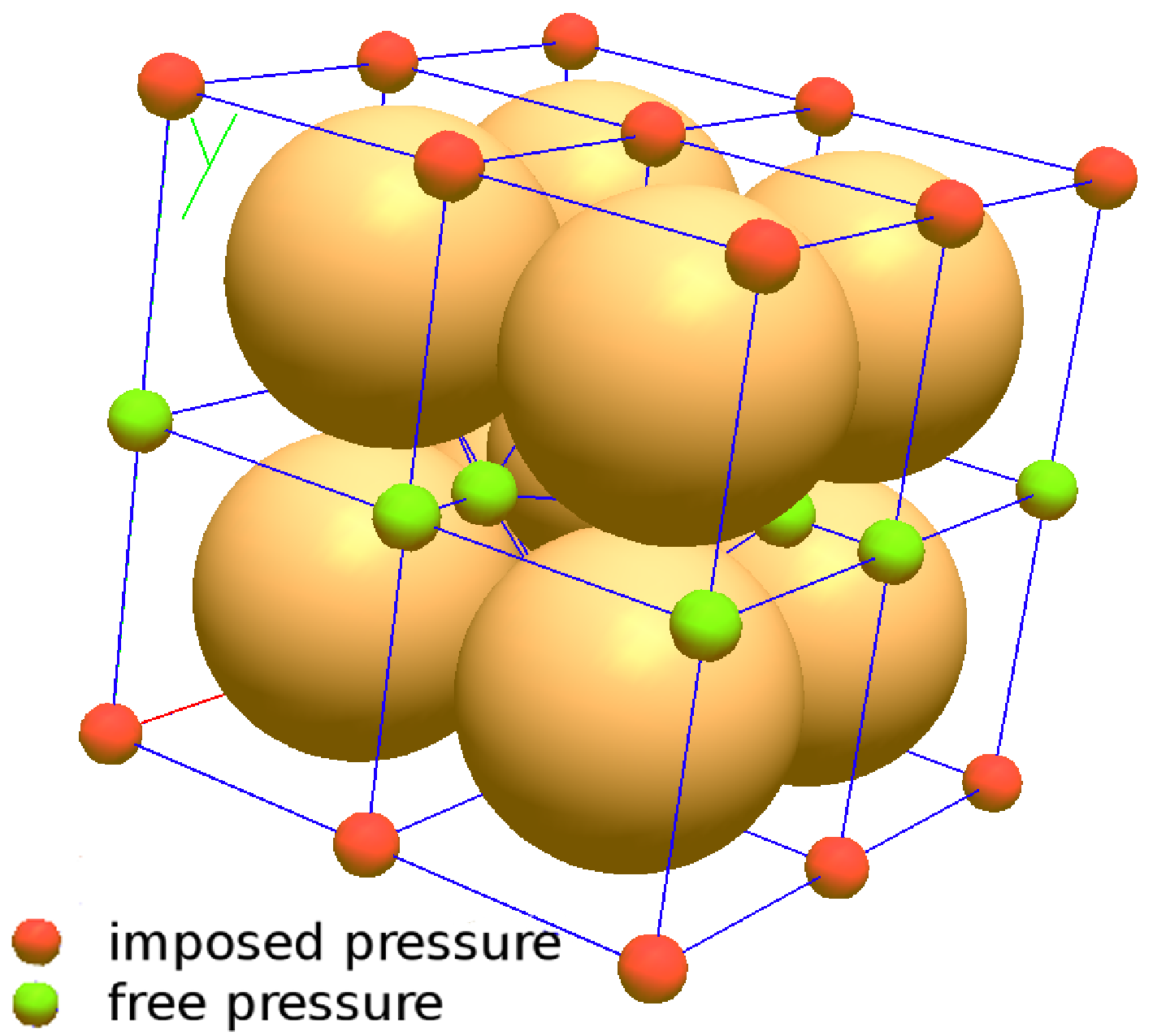}{(c)}
\caption{Boundary conditions of flow simulations (a), FEM mesh (b) and FV mesh (c) of a 9-spheres packing. The FV
packings are plotted with Voronoi graph, where pressure values are defined at each point.}
\label{fig:boundary_conditions}
\end{figure}

\begin{figure}
\centering
\includegraphics[width=50mm]{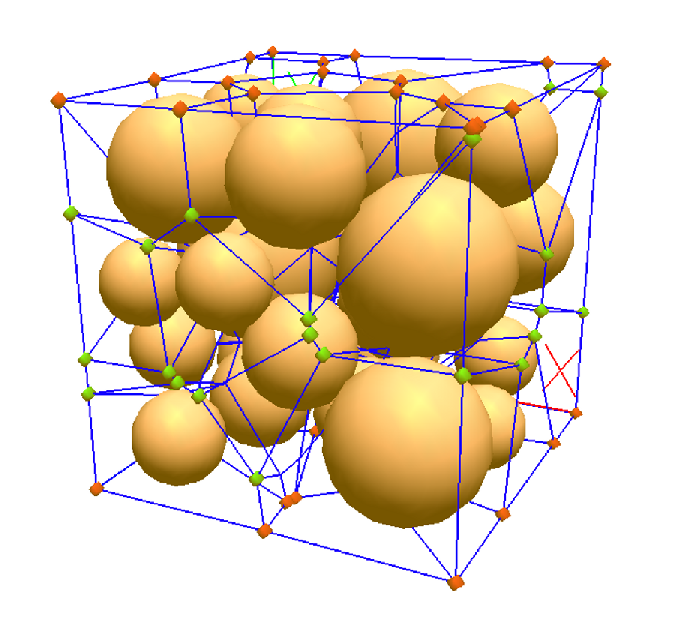}{(a)}
\includegraphics[width=45mm]{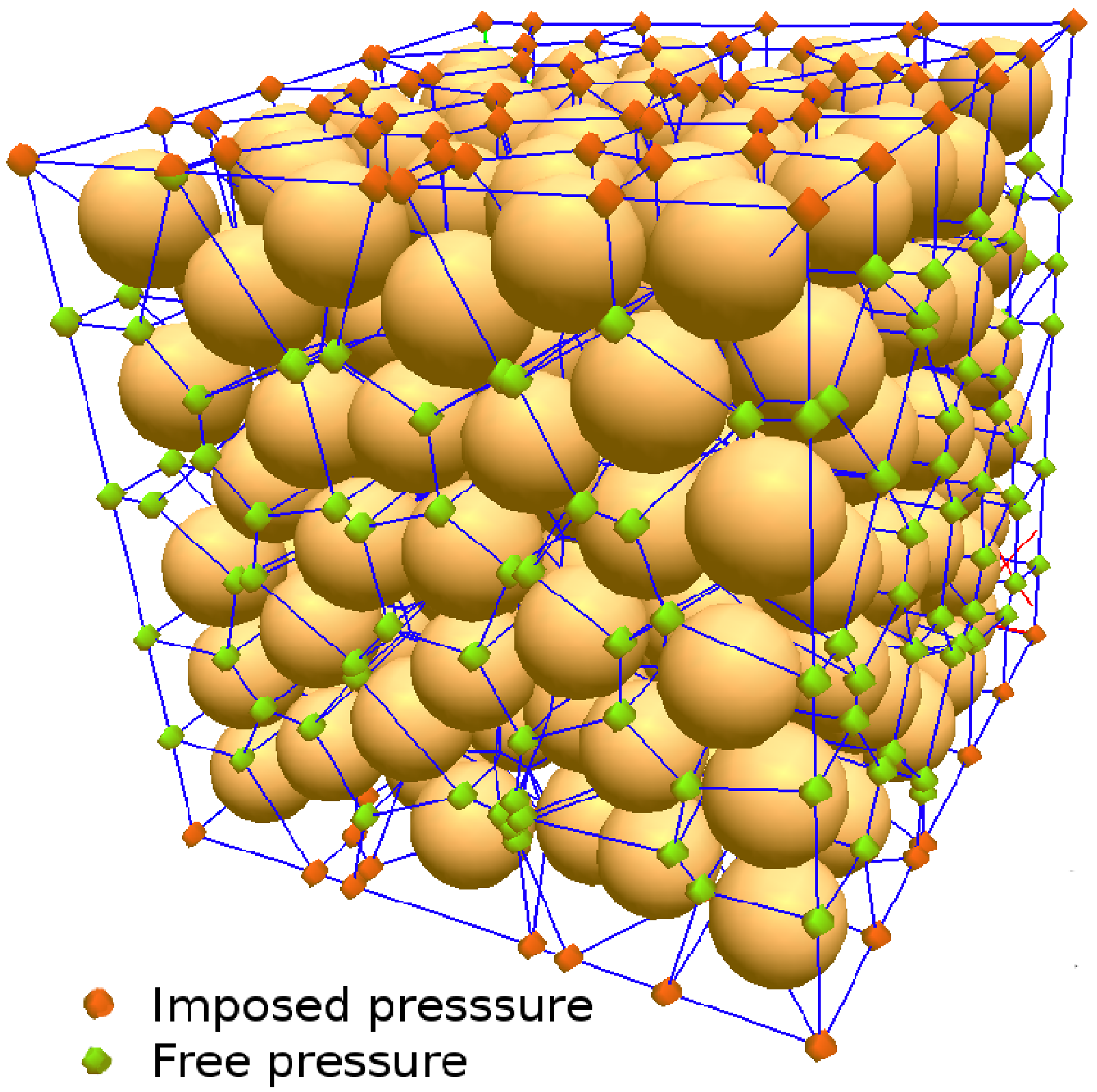}{(b)}
\caption{Packings of 25 and 200 spheres with corresponding Voronoi graph.}
\label{fig:packings}
\end{figure}

The smallest test problem we consider is a regular cubic packing of 8 spheres. Packings of nine spheres are obtained by
placing an additional sphere (with variable size $d$) in the center of the cubic packing
(fig.~\ref{fig:boundary_conditions}). Larger assemblies of 25, 54, and 200 spheres (fig. \ref{fig:packings}) are random
packings that were
generated using the DEM software Yade (\citealt{YadeDoc}) by simulating the growth of spheres after random positioning,
using the algorithm of Chareyre et al. \citeyearpar{Chareyre2002}. Radii are generated randomly with respect to uniform
distributions between radii $r_{min}$ and $r_{max}$. In the 200-spheres assembly, sizes are narrow graded, with
$r_{min}/r_{max}=0.9$ (this small dispersion of sizes prevents the formation of crystal-like paterns). Sizes in the
25-spheres and 54-spheres packings, are more dispersed, with $r_{min}/r_{max}=0.5$. After dense and stable packings are
obtained, the spheres's positions are fixed.

For each packing geometry, the flow boundary conditions are the ones described on fig.~\ref{fig:boundary_conditions}:
pressure is imposed on the top and bottom boundaries, while no-flux conditions are imposed on the lateral boundaries.

Both ``no-slip'' and ``symmetry'' conditions are considered on the lateral boundaries. In the FV modeling, they are
reflected in the definition of conductivity and forces in the throats $\Theta_{ij}$ in contact with boundaries as
follows:
\begin{itemize}
\item For the ``no-slip'' condition,  the surfaces of infinite spheres in contact with
$\Theta_{ij}$ are accounted for in Eqs. \ref{eq:hydraulicradius} and \ref{eq:dragforcek}, as for any other sphere.
\item For the ``symmetry'' condition, these surfaces are ignored : $\gamma^k_{ij}=0$, when $k$
is the indice of a bounding sphere.
\end{itemize}

Table \ref{tab:problemsize} gives a comparison of problems sizes in terms of degrees of freedom (DOF) of the pressure
field,
and CPU times for solving. The data in this table is only an indication of how much is gained from the pore-scale
formulation of Stokes
flow: impacts of coarsening the FEM mesh, or benefits of other algorithms in the FV model (e.g. conjugated gradient)
have not been investigated yet.

\begin{table}
\caption{\label{tab:problemsize} Compared DOF's and CPU time. (a) includes only solving Stokes flow, excludes
preprocessing (e.g. mesh generation) and postprocessing (forces on spheres) done via a graphical user interface; (b)
includes packing triangulation, solving, and computation forces on particles; (c) out-of-memory: no result.}
\begin{tabular}{@{}p{9mm}@{\ \ \ \ }l@{\ \ }l@{\ \ }p{17mm}@{\ \ }p{15mm}}
\hline\noalign{\smallskip}
Nb. of spheres & FEM dof's & FV dof's & FEM time$^{(a)}$ (s) & FV time$^{(b)}$ (s) \\
\hline\noalign{\smallskip}
9 & 1.7e5 & 45 &  300& 0.00022 \\
200 & 1.2e6 & 1093 &  5400& 0.0046\\
2e3 & ${(c)}$ & 11.6e3 & ${(c)}$ & 0.091\\
2e4 & ${(c)}$ & 11.3e4 & ${(c)}$ & 2.21\\
\hline\noalign{\smallskip}
\end{tabular}
\end{table}
%
\subsection{Pressure Field}
\begin{figure}
\centering
\includegraphics[width=55mm]{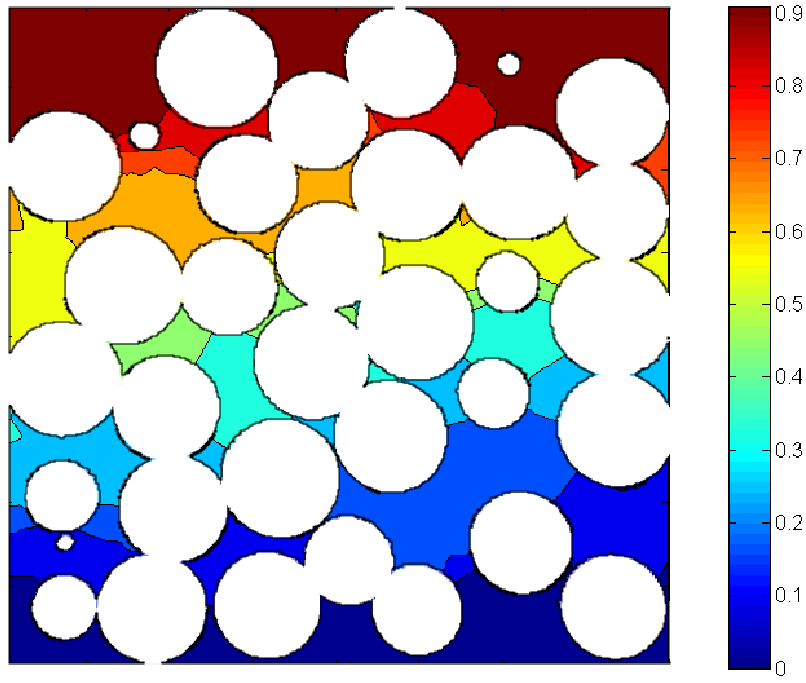}{(a)}
\includegraphics[width=55mm]{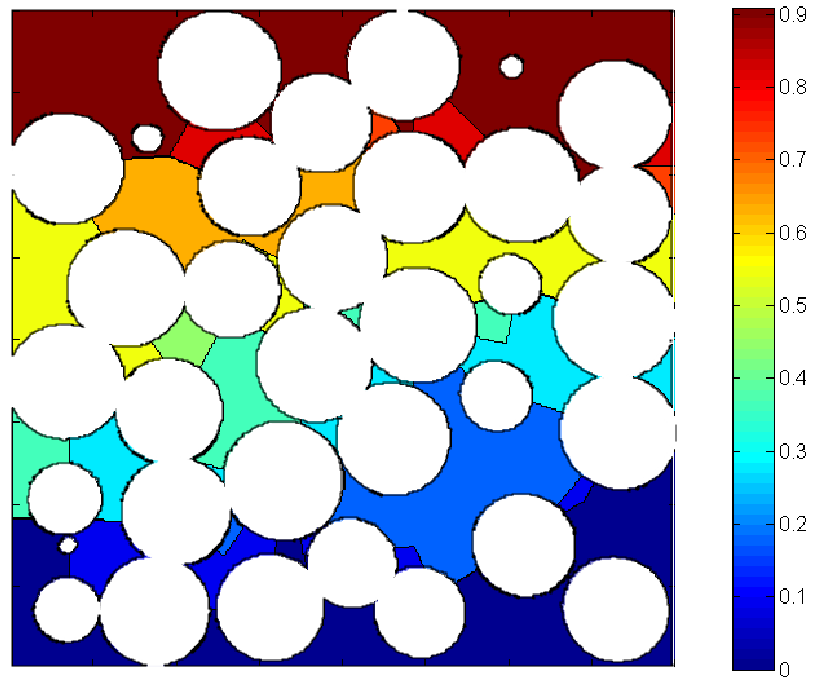}{(b)}
\caption{Isovalues of pressure in the 200 spheres packing on the plane $x=0.5 l_0$, obtained with FEM (a) and FV (b).}
\label{fig:pressure200}
\end{figure}

The pressure fields from FEM and FV are compared on figure \ref{fig:pressure200}.  Local conductances are defined as in
eqs. (\ref{eq:poiseuillecylinder}-\ref{eq:hydraulicradius}). The isovalues in the FV result are, by model definition
(element-wise constant pressure), coincident with
facets. Thus, the curvature of some isolines, as observed in FEM results, cannot be reproduced.
The curvature of the isolines in the FEM solution is, however, usually very small so that most of them are
reasonably approximated by straight lines. Furthermore, most isolines are found near the necks of
the flow paths, thus justifying a posteriori the approximation of element-wise constant pressure used in the
FV scheme. Overall, the two fields compare well. The deviation from the horizontal isolines that would be obtained in an
homogeneous continuum with same boundary conditions is relatively well reproduced by the FV model. 

\subsection{Fluxes}
Two fluxes can actually be obtained from simulations: influx $Q_i$, and outflux $Q_o$ at the respective boundaries (in
the FV model, they are computed by summing the $q_{ij}$ of eq. \ref{eq:poiseuillecylinder} for elements where pressure
is imposed). It
was found that the difference between $Q_i$ and $Q_o$ is always negligible (less than $10^{-6} \times |Q_i|$). This is
consistent with the incompressibility assumption and indicates a good convergence of the numerical solvers, be it in FEM
(direct sparse solver Pardiso) or FV (Gauss Siedel method). In the following, we define $Q=Q_i$.
Fluxes are expressed in the form of normalized permeability defined as $K=\mu \, Q \, h_0/(\Delta P S^2)$,
with $S=l_0^2$ the packing's cross-sectional area. Note that the introduction of permeability is only for the
convenience of
results representation, and is not implying anything with respect to the \textit{representative volume element} (REV)
concept. The topic of this paper is the comparison of small-scale modelings of fluxes and forces
for some spheres packings: derivation of higher-scale properties of an equivalent continuum is beside the scope of the
present investigation.

\begin{figure}
\centering
\includegraphics[width=85mm]{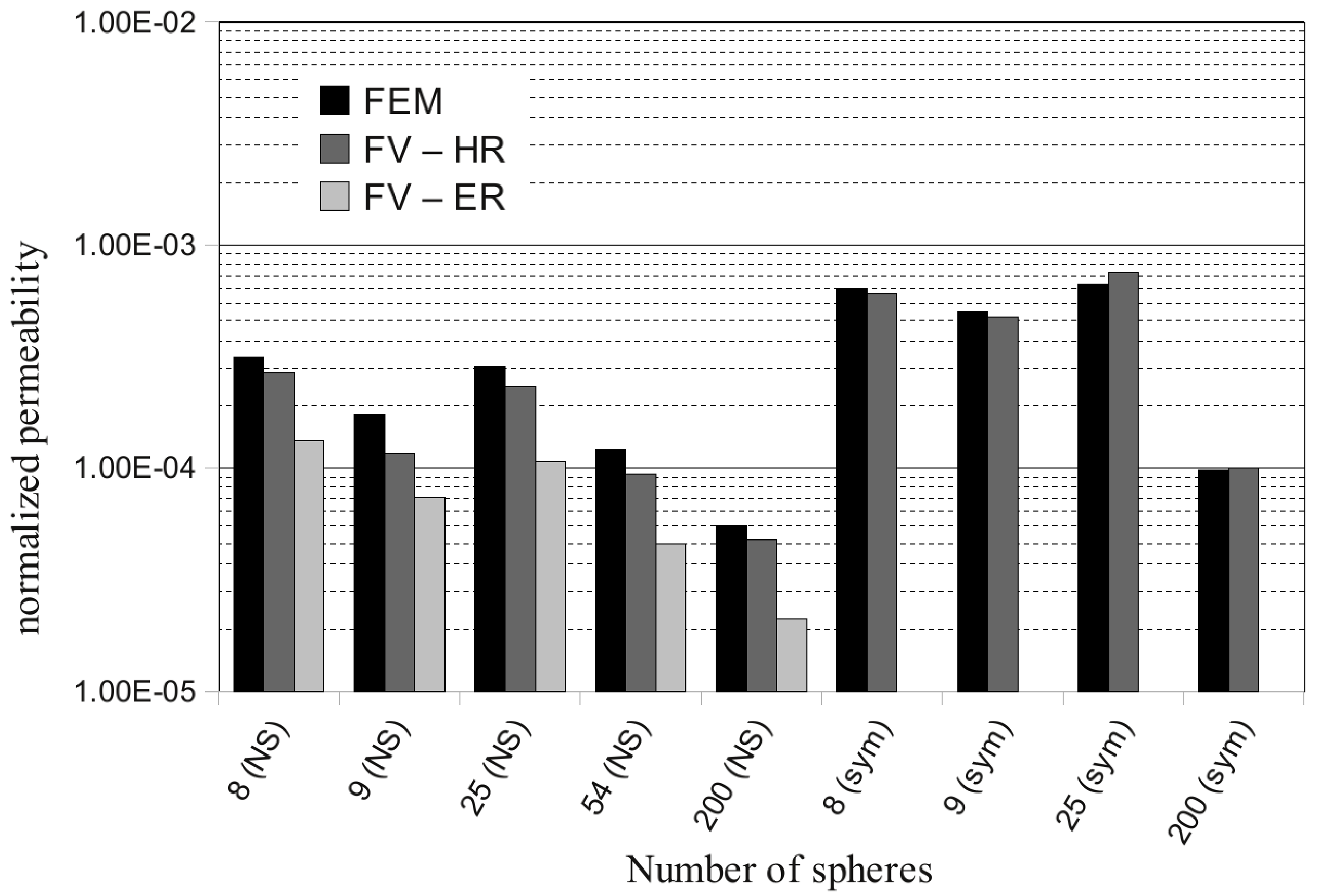}
\caption{Predicted permeability in FEM and FV versus size of packing, for no-slip (NS) and symmetric (sym) boundary
conditions. The FV results include conductances defined using the hydraulic radius (HR - eq. \ref{eq:hydraulicradius})
and effective radius (ER - eq. \ref{eq:Reffg}).}
\label{fig:perm_vs_n}
\end{figure}

Permeabilities obtained with FEM and FV for the different packings are compared on fig. \ref{fig:perm_vs_n}.
Both no-slip and symmetry conditions are considered for lateral boundaries. Comparison between the definitions of
conductivity using the ``hydraulic`` radius (HR) of eq. \ref{eq:hydraulicradius} and Bryant's ''effective`` radius (ER)
of eq. \ref{eq:Reffg} are provided. This later comparison only covers no-slip conditions, as the definition of effective
radii in the symmetry case was ambiguous.

Symmetric boundary conditions give an average ratio of 1.01 between $K_{HR}$ and $K_{FEM}$, with a maximum deviation of
$+13\% $ for the 25-sphere packing.
With no-slip conditions, the ratio are 0.78 for $K_{HR}/K_{FEM}$ (max. deviation $-40\% $) and 0.39 for $K_{HR}/K_{FEM}$
(max. deviation $-63\% $);
$K_{HR}$ is giving the best estimate of $K_{FEM}$ in all cases. The fact that the FEM results are better reproduced for
symmetric conditions suggests that the fluxes along planes are underestimated. We consider, however, that the
predictions of fluxes from hydraulic radius can be considered satisfactory overall.

\begin{figure}
\centering
\includegraphics[width=55mm,angle=-90]{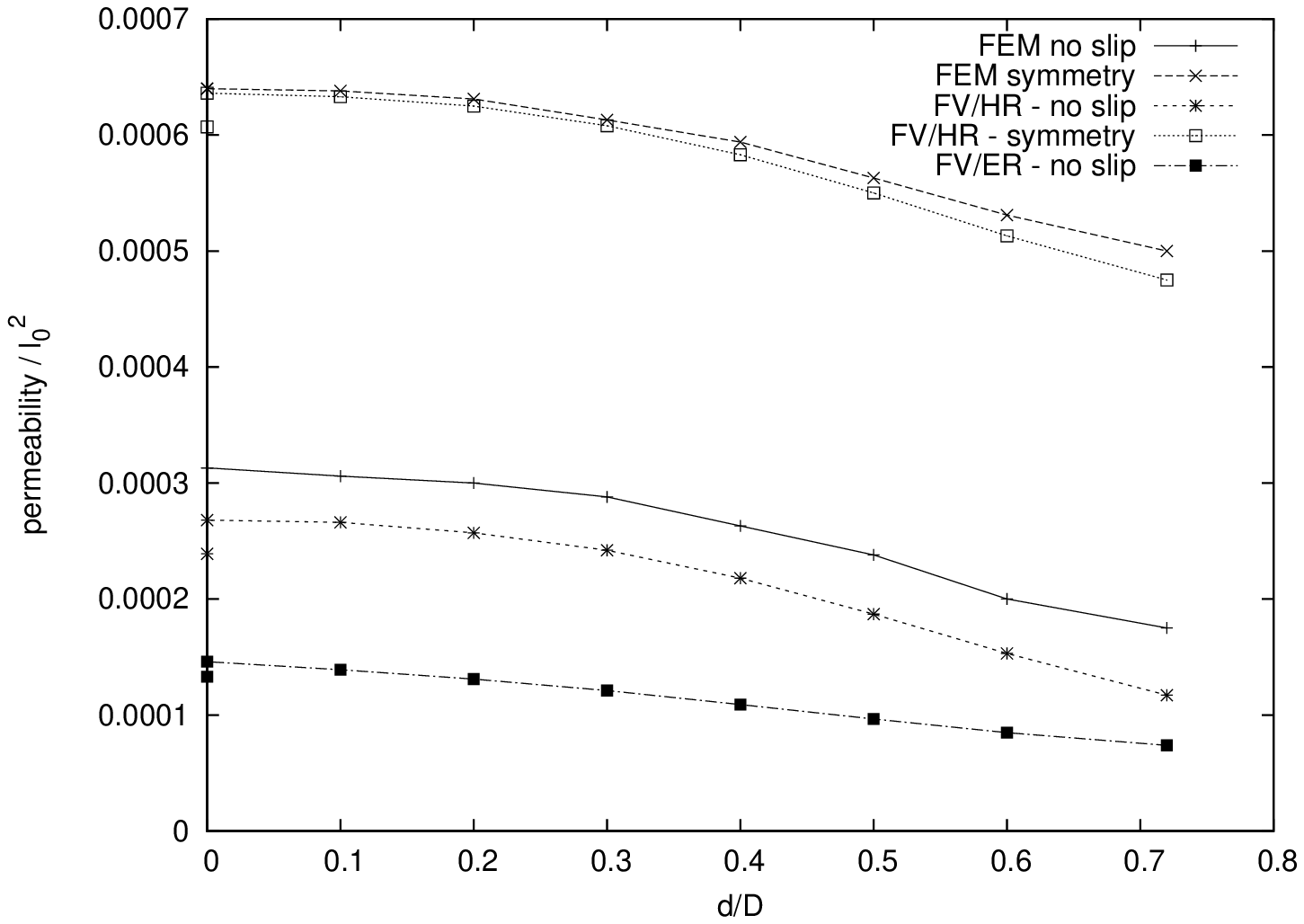}
\caption{Permeability of 9 spheres packing versus size of the the central particle.}
\label{fig:perm9}
\end{figure}

The evolution of permeability in 9-spheres packings as a function of size $d$ of the inner sphere's size (fig.
\ref{fig:boundary_conditions}) is plotted on fig. \ref{fig:perm9}. The evolution of $K$ with $d/D$ is correctly reflected in HR-based results.
Again, FV results match FEM better for symmetric boundaries. ER-based simulations are
underestimating the fluxes by an average factor $K_{ER}/K_{FEM}=0.43$.

It can be concluded that the initial value $\alpha=\nicefrac{1}{2}$ of the conductance factor entering eq.
\ref{eq:poiseuillecylinder} tends to underestimate fluxes in average. Although systematic comparisons could enable a
better calibration of $\alpha$, deviation from the FEM results is an inherent defect due to the approximations of the
pore-scale description adopted here. Considering the small number of samples we tested, $\alpha$ has not been further
adjusted to closely match FEM results.

\begin{figure}
\centering
\includegraphics[width=80mm]{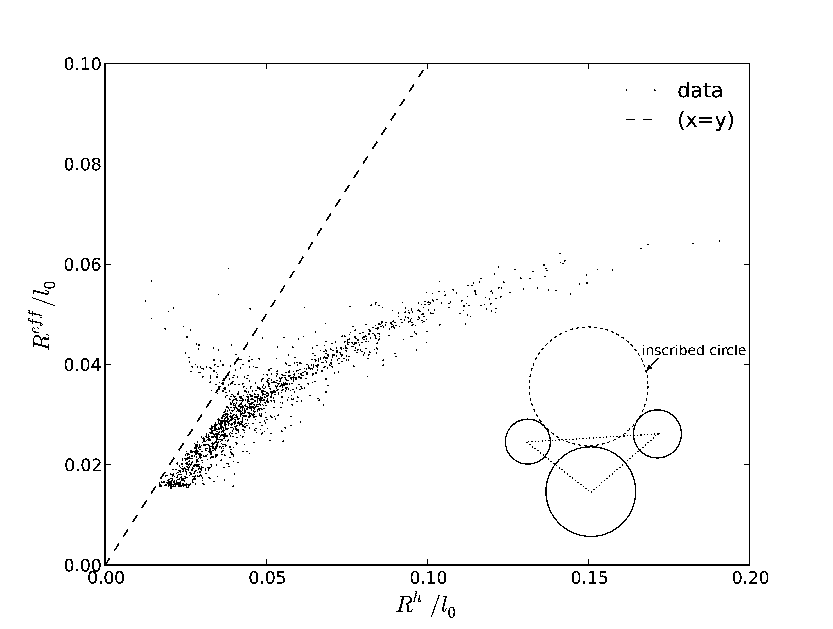}
\caption{Effective radius versus hydraulic radius for all facets in the 200 grains packing (a), with exemple of a facet
giving $R_{ij}^{eff}>R_{ij}^{h}$.}
\label{fig:rh_vs_reff}
\end{figure}

To investigate further the differences found between $K_{HR}$ and $K_{ER}$, the values of $R^h$ and $R^{eff}$ obtained
for all facets of the 200 spheres packing triangulation have been plotted (fig \ref{fig:rh_vs_reff}). Clearly, the two
values are strongly correlated, with the exception of very few points that will be commented below, but they deviate
from the $x=y$ line. The average value of $R^h$ and $R^{eff}$ is 1.5. Considering this distribution, it is obvious that
the differences found between $K_{HR}$ and $K_{ER}$ are reflecting the fact that $R^h>R^{eff}$ in average at the local
scale.

A small number of points fall out the well correlated cloud, for which $R^{eff}>R^{h}$. A close inspection of the outliers revealed that these correspond to corner cases similar to the one represented on fig.~\ref{fig:rh_vs_reff}, where a flat triangle results in the inscribed circle overlapping outside its original facet. In such a situation, a flatter triangle will result in a smaller hydraulic radius (smaller $\phi_{ij}$), but a higher effective radius. 
The inscribed circle will eventually overlap other inscribed circles or spheres of the
packing, thus loosing physical consistency: it was not clear to us how such special cases should be handled. The number of affected facets is, however, small and do not significantly affect the results. These results lead us to conclude that he hydraulic radius is overall a more robust parameter than the effective radius.
%
\subsection{Forces}
Forces on particles and boundaries, as defined in section \ref{sec:forces_formulation}, have been obtained for all
packings using the FV scheme. In all cases, the sum of pressure and viscous forces on spheres and
boundaries is, as expected, close to $\Delta P \times S$ ($\pm 10^{-5}\%$), where $S$ is the packing's cross-sectional
area. Comparison for the  $8$-spheres packing is trivial: $F_i=S \times \Delta P/8$ on each sphere for both FEM and FV, modulo numerical errors. A detailed comparison is presented here only for the $9$-spheres packing, for which  contour integral of fluid stress in the FEM models (the FEM results in table \ref{tab:forces} have been collected.

Table \ref{tab:forces} gives the total fluid forces per element of the solid phase, and details the viscous and pressure
contributions. The viscous
force on lateral boundaries, total force on big spheres, and total force on the center sphere, are all approximated with
an error smaller that 10\%. Surprisingly, the prediction is less good for individual viscous and pressure contributions
than for
the total force. The viscous part is overestimated, while the pressure effect is underestimated. The two errors
compensation
results in a smaller error on the total force.

\begin{figure}
\centering
\includegraphics[width=60mm,angle=-90]{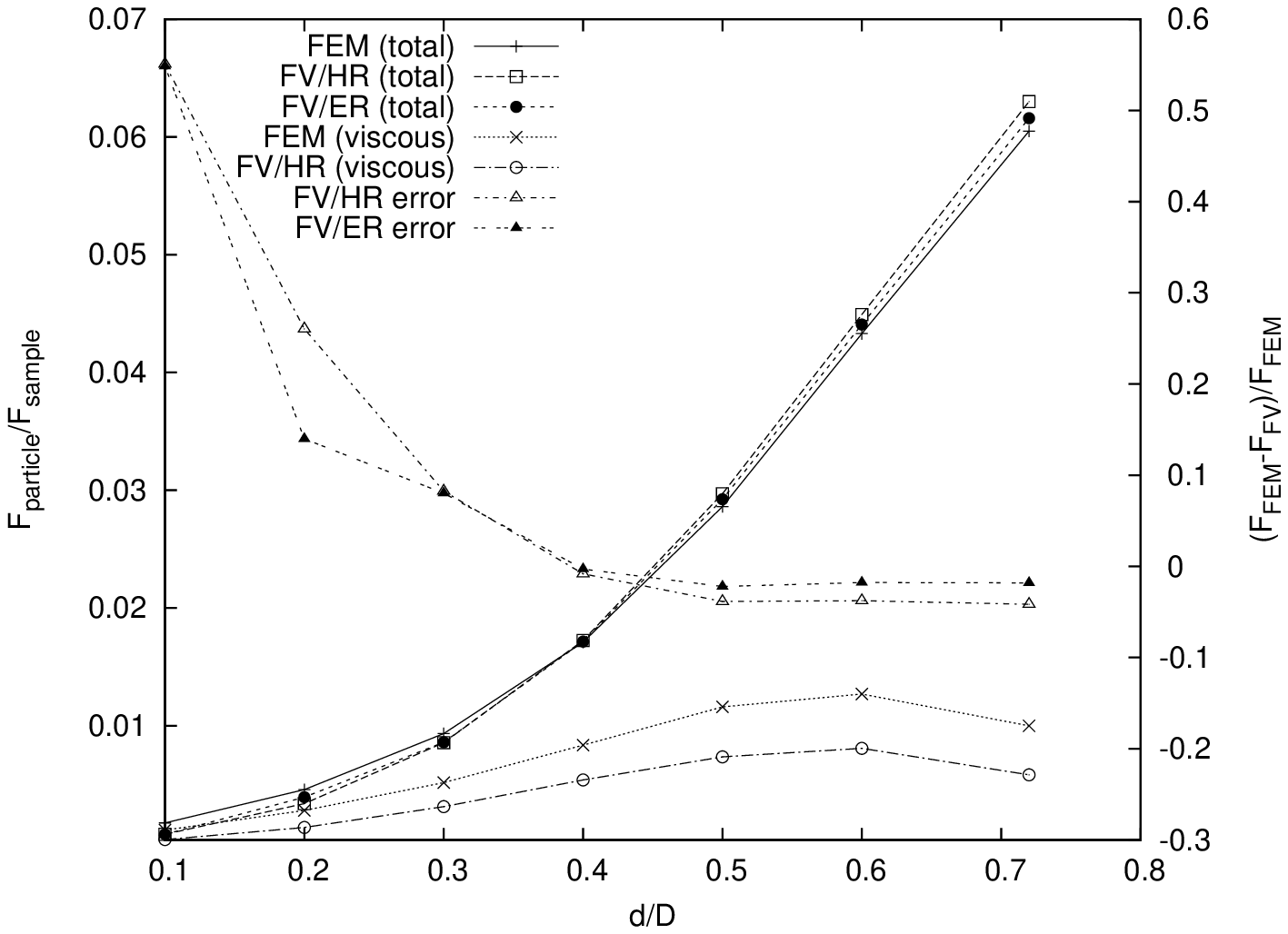}
\caption{Total force and viscous force applied on a particle of size $d$, placed in the center of a cubic packing of 8
particles of size $D$, in FEM simulations, FV simulations based on hydraulic radius (HR), and FV simulations based on
effective radius (ER).}
\label{fig:forces_small}
\end{figure}

The evolution of force terms and error for the center sphere is given on fig. \ref{fig:forces_small}. This result shows
an increasing error with decreasing $d/D$. For $d/D>0.1$, the FV scheme underestimates the force on the the center
sphere by a factor 2. This suggest a limitation of the current formulation in the limit of (1) high
contrasts in particles size, and/or (2) small particles ``floating'' in the voids between big ones. Here again, more
investigations are needed in order to determine what situation exactly is generating this deviation from the FEM
solution. As long as $d/D>0.2$, though the predicted force is very well predicted, with error less than 10\%. It is
observed that similar forces are obtained with hydraulic radius (HR) and effective radius (ER).

\begin{table}
\caption{\label{tab:forces} Normalized forces in 9 spheres packings ($d/D=0.72$). All forces are divided by $S.\Delta
P$, so that the sum of all forces should be exactly 1.}
\begin{tabular}{lll}
\hline\noalign{\smallskip}
& FEM & FV\\
\noalign{\smallskip}\hline
\multicolumn{3}{l}{\textit{Forces on one big sphere (y-component)}}\\
viscous force & $1.51\times 10^{-2}$ & $1.06\times 10^{-2}$ \\
pressure force & $9.09\times 10^{-2}$ & $9.64\times 10^{-2}$ \\
total & $1.06\times10^{-1}$ & $1.07\times10^{-1}$\\
\noalign{\smallskip}\hline
\multicolumn{3}{l}{\textit{Forces on the small sphere (y-component)}}\\
viscous force & $9.38\times10^{-3}$ & $5.83\times10^{-3}$ \\
pressure force & $5.17\times10^{-2}$ & $5.72\times10^{-2}$ \\
total & $6.04\times10^{-2}$ & $6.30\times10^{-2}$ \\
\noalign{\smallskip}\hline
\multicolumn{3}{l}{\textit{Forces on boundary x=0}}\\
viscous force (y-component) & $2.27\times10^{-2}$ & $2.03\times10^{-2}$ \\
pressure force (x-component) & $4.98\times10^-1$ & $5.00\times10^-1$ \\
\noalign{\smallskip}\hline
\multicolumn{3}{l}{\textit{Total force on the solid phase }}\\
y-component & $1-5.56\times10^{-4}$& $1+1.18\times10^{-4}$\\
\hline\noalign{\smallskip}
\end{tabular}
\end{table}

\section{Conclusion}
The pore-space of a dense spheres packing can be efficiently represented by means of a regular Delaunay triangulation.
In this way, the total number of DOFs and CPU time required for calculating flow and forces acting on the particles are
reduced drastically with respect to small scale Stokes flow FEM calculations.

Expressions of local conductivities and forces induced on the particles have been derived. Bounding planes are
straightforwardly taken into account in the method,
and their effect in terms of permeability and viscous forces is reflected in the model for both no-slip and symmetry
conditions. The method could be extended to
account for non non-spherical particle shapes, as approximated by the intersection of a finite number spheres of various
diameter.

The definition of local conductivity of pore throats is based on a definition of the local \textit{hydraulic radius}
$R^h$, as in the Kozeny-Carman (KC) derivation of permeability in porous material. The difference with KC-like
derivations lies in the fact that connectivity and tortuosity of the network do not have to be included in the
derivation, since they arise directly from the tetrahedral domain decomposition.
The definition of the hydraulic radius $R^h$ is simple and involves almost only real and 3D vectors multiplications for
the definition of
surfaces and volumes, which favors numerical performance.

Taking the FEM results as reference, the permeabilities obtained with the pore-scale modeling are in most cases within
the range of $\pm 20\%$, usually below the reference value. Considering only symmetric boundary conditions, the
permeabilites are better predicted (in the range $\pm 10\%$), which indicates that fluxes along planar no-slip
boundaries are underestimated. We consider that such estimates are acceptable overall, keeping in mind that permeability
prediction is of secondary importance, in comparison with forces, for the hydromechanical couplings.

The effective radius proposed by Bryant et al. \citeyearpar{Bryant1993} for mono-sized spheres also gives acceptable predictions of permeabilities for polydispersed sphere packings, but it is computationally more expensive and underestimates fluid fluxes more than the hydraulic radius proposed in this investigation. It is important to note that the pore network model of Bryant et al. includes a reduction of $L_{ij}$, accounting for
overlapping
cylindrical throats. This correction has not been implemented because it needs relatively complex computations of
cylinders intersections. By maximizing micro-gradients, this correction could have at least partly balanced the
underestimation of permeabilities in our results. Hence, our findings cannot be considered in disagreement with Bryant
et al. model in itself. They only indicate that the effective radius should not be used in connection with the regular
Delaunay partitioning we are developing.

The pressure field compares well with
the FEM results. In all cases, the sum of forces applied on the solid phase in a unit cube under unit gradient is close
to
one, showing the correct implementation and validity of the model at the mesoscale. Forces applied by the fluid on
individual particles, and viscous forces on no-slip boundaries,
are found to be correct, with errors typically less than 10\%: the error, though, grows larger when the size ratio $d/D$
of the particles is less than 0.2.

This modeling of fluid forces gives a sound basis for fluid-particles systems. Current works focus on the fully coupled
problems in deformable packings with deformation-induced pore pressure, and on the optimization of the flow computation in this context.

\begin{acknowledgements}
This work and PhD grant of E. Catalano is supported by Grenoble Institute of Technology through BQR-2008 program.
A. Cortis' work was supported, in part, by the U.S. Department of Energy under Contract N0. DE-AC02-05CH11231.
\end{acknowledgements}

\end{document}